\documentclass[journal=jpclcd,manuscript=letter,layout=twocolumn]{achemso}
\usepackage{amsmath,graphics,epsfig,color,verbatim,gensymb}
\usepackage[T1]{fontenc}
\usepackage[utf8]{inputenc}
\usepackage{amssymb}
\DeclareUnicodeCharacter{0146}{\c{n}}

\title{Inverse Design of Ultralow Lattice Thermal Conductivity Materials Via Lone Pair Cation Coordination Environment}
\author{Eric B. Isaacs}
\affiliation{Department of Materials Science and Engineering, Northwestern University, Evanston, Illinois 60208, USA}
\author{Grace M. Lu}
\affiliation{Department of Materials Science and Engineering, Northwestern University, Evanston, Illinois 60208, USA}
\alsoaffiliation{Present Address: Department of Materials Science and Engineering, University of Illinois at Urbana-Champaign, Urbana, IL 61801, USA}
\author{Christopher Wolverton}
\email{c-wolverton@northwestern.edu}
\affiliation{Department of Materials Science and Engineering, Northwestern University, Evanston, Illinois 60208, USA}

\begin{tocentry}
  \includegraphics[width=3.25in]{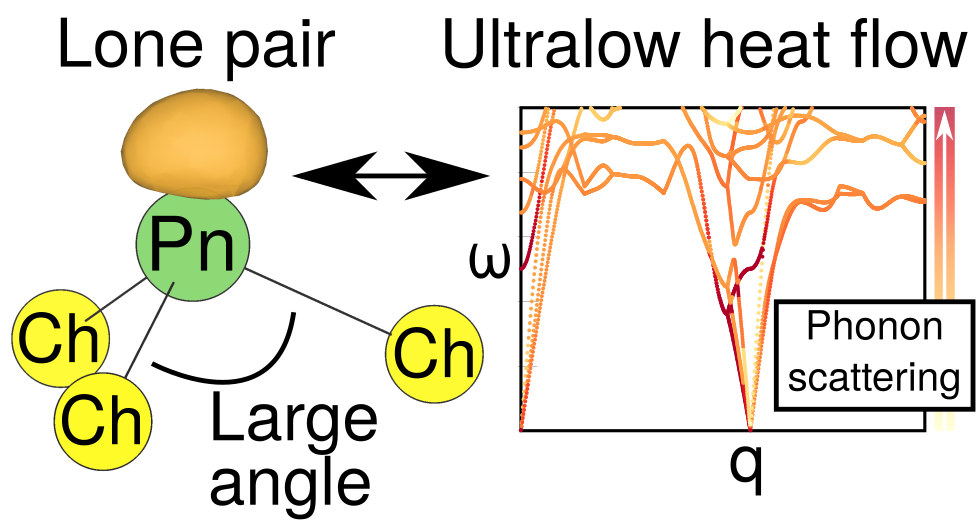}
\end{tocentry}

\begin{document}

\begin{abstract}
  The presence of lone pair (LP) electrons is strongly associated with
  the disruption of lattice heat transport, which is a critical
  component of strategies to achieve efficient thermoelectric energy
  conversion. By exploiting an empirical relationship between lattice
  thermal conductivity $\kappa_L$ and the bond angles of pnictogen
  group LP cation coordination environments, we develop an inverse
  design strategy based on a materials database screening to identify
  chalcogenide materials with ultralow $\kappa_L$ for thermoelectrics.
  Screening the $\sim$ 635,000 real and hypothetical inorganic
  crystals of the Open Quantum Materials Database based on the
  constituent elements, nominal electron counting, LP cation
  coordination environment, and synthesizability, we identify 189
  compounds expected to exhibit ultralow $\kappa_L$. As a validation,
  we explicitly compute the lattice dynamical properties of two of the
  compounds (Cu$_2$AgBiPbS$_4$ and MnTl$_2$As$_2$S$_5$) using
  first-principles calculations and successfully find both achieve
  ultralow $\kappa_L$ values at room temperature of $\sim$ 0.3--0.4
  W/(m$\cdot$K) corresponding to the amorphous limit. Our data-driven
  approach provides promising candidates for thermoelectric materials
  and opens new avenues for the design of phononic properties of
  materials.
\end{abstract}


Minimizing lattice thermal transport is critical for thermoelectric
heat-to-electricity conversion, in which such heat flow acts as a loss
mechanism.\cite{goldsmid_introduction_2010} This can be seen in the
dimensionless thermoelectric figure of merit to be maximized
  $$ZT = \frac{\sigma S^2}{\kappa_L + \kappa_e}T,$$ in which the
  lattice thermal conductivity $\kappa_L$ appears in the denominator.
  Here, $\sigma$ is electrical conductivity, $\kappa_e$ is electronic
  thermal conductivity (typically small in semiconductors), $S$ is
  thermopower, and $T$ is temperature. Achieving ultralow values of
  $\kappa_L$ [i.e., $\lesssim 1$ W/(m$\cdot$K)] is an integral
  component of successful strategies to design high-performance
  thermoelectric materials.\cite{snyder_complex_2008}


  While there exist various strategies to extrinsically reduce
  $\kappa_L$, such as nanostructuring, alloying, and doping (see,
  e.g., Refs.
  \citenum{majumdar_thermoelectricity_2004,poudel_high-thermoelectric_2008,minnich_bulk_2009,vineis_nanostructured_2010,garg_role_2011,biswas_high-performance_2012,tian_phonon_2012,hori_phonon_2013,shiga_influence_2014,lindsay_survey_2018,hao_computational_2019}),
  a conceptually appealing alternative approach is to instead utilize
  materials with intrinsically ultralow
  $\kappa_L$.\cite{jana_crystalline_2018} One characteristic related
  to ultralow $\kappa_L$ in certain materials is the presence of a
  lone pair (LP), a pair of valence electrons localized on a single
  atom.\cite{iupac_lone_pair} A LP is typically achieved via a cation
  (called a LP cation) whose two valence $s$ electrons remain
  localized on or near the cation. For example, given Sn's electronic
  configuration of [Kr] $4d^{10}$\ $5s^2$\ $5p^2$, Sn$^{2+}$ (as in
  SnSe) is a LP cation, whereas Sn$^{4+}$ (as in SnSe$_2$) is not.
  Elements commonly forming LP cations include elements in group III
  (Tl; 1+ oxidation state), group IV (Ge, Sn, Pb; 2+ oxidation state),
  and the pnictogen group (P, As, Sb, Bi; 3+ oxidation state) of the
  periodic
  table.\cite{greenwood_chemistry_1997,seshadri_visualizing_2001,waghmare_first-principles_2003,stoltzfus_structure_2007,walsh_stereochemistry_2011}



  It has been long known that the presence of LP cations is related to
  low thermal conductivity. In 1961, Petrov and Shtrum observed that
  thermal conductivity values of
  A$^{\mathrm{I}}$B$^{\mathrm{V}}$X$_2^{\mathrm{VI}}$ compounds (whose
  B$^{3+}$ is a LP cation) are significantly lower than those of
  A$^{\mathrm{I}}$B$^{\mathrm{III}}$X$_2^{\mathrm{VI}}$ compounds
  (whose B$^{3+}$ is not a LP cation), where A$^{\mathrm{I}}$,
  B$^{\mathrm{III}}$, and B$^{\mathrm{V}}$ are monovalent, trivalent,
  and pentavalent elements, respectively, and X$^{\mathrm{VI}}$ is a
  chalcogen.\cite{petrov_heat_1962}. For example, room-temperature
  $\kappa_L$ for AgSbSe$_2$ is only 0.7 W/(m$\cdot$K), as compared to
  1.8 W/(m$\cdot$K) for AgInSe$_2$.\cite{morelli_intrinsically_2008}
  Similarly, room-temperature $\kappa_L$ for Cu$_3$SbSe$_4$ (whose
  Sb$^{5+}$ is not a LP cation) is 2.5--3.5 W/(m$\cdot$K), much larger
  than the corresponding 0.7--1.0 W/(m$\cdot$K) for Cu$_3$SbSe$_3$
  (whose Sb$^{3+}$ is a LP cation), despite the two compounds
  containing the same elements and having similar
  stoichiometry.\cite{skoug_structural_2010,zhang_first-principles_2012}
  Therefore, it is not surprising that several of the most promising
  thermoelectrics (e.g., PbTe, Bi$_2$Te$_3$, and SnSe) contain LP
  cations and that many other compounds containing LP cations have
  also been under
  investigation.\cite{mcguire_exploring_2005,dutta_bonding_2019,mukhopadhyay_two-channel_2018,dong_bournonite_2015,zhao_bicuseo_2014,du_impact_2017,chetty_tetrahedrites_2015,zhang_screening_2017}

  Empirically, the coordination environment of a LP cation was found
  by Skoug and Morelli to be intimately related to lattice thermal
  transport.\cite{skoug_role_2011} In particular, within a set of
  compounds containing pnictogen group lone pair cations (L)
  coordinated to a given number of chalcogens (X), e.g., compounds
  whose L have 3 nearest neighbor X atoms, they observed a strong
  negative correlation between measured room-temperature $\kappa_L$
  and a local structural parameter related to the bond angles
  $\angle$X--L--X of the LP cation's coordination cage. This parameter
  $\overline{\alpha}^{(s)}$ can be defined as
$$\overline{\alpha}^{(s)} = \frac{1}{N_{\mathrm{ang}}^{(s)}} \sum_i \alpha_i^{(s)}
\theta(\alpha_{\mathrm{max}} - \alpha_i^{(s)}),$$ where
$\alpha_i^{(s)}$ is the $i^{\mathrm{th}}$ $\angle$X--L--X for the
$s^{\mathrm{th}}$ LP cation site, $\theta$ is the Heaviside step
function, $\alpha_{\mathrm{max}}$ is
$\cos^{-1}(-\frac{1}{3}) \approx 109.5\degree$ (bond angle for ideal
tetrahedral coordination), and
$N_{\mathrm{ang}}^{(s)} = \sum_i \theta(\alpha_{\mathrm{max}} -
\alpha_i^{(s)}).$ In words, $\overline{\alpha}^{(s)}$ is simply the
average of the $N_{\mathrm{ang}}^{(s)}$ bond angles no larger than
109.5$\degree$ for the $s^{\mathrm{th}}$ LP cation site. The quantity
$\overline{\alpha}^{(s)}$, which originates from a study of Sb
chalcogenides using the bond valence sum concept, was found to relate
to the effective valence of the LP cation (i.e., between the nominal
3+ and 5+ common oxidation states for Sb) and was interpreted as
describing the retraction of the LP from the LP
cation.\cite{wang_studies_1996} In general, the work of Skoug and
Morelli (and others\cite{wang_lone-pair_2018}) strongly suggests the
importance of the LP cation environment and spatial distribution of
the LP in influencing lattice dynamics.

In this Letter, we exploit the connection between LP cation
coordination environment and lattice heat transport with the goal of
designing ultralow $\kappa_L$ materials. We develop an inverse design
approach based on a materials database screening, which can be
considered the so-called ``second modality'' of inverse
design\cite{zunger_inverse_2018} and is related to our previous works
on inverse design for electronic band
structure.\cite{isaacs_inverse_2018,isaacs_materials_2019} Searching
for synthesizable compounds containing pnictogen group LP cations
coordinated to chalcogens with large $\overline{\alpha}^{(s)}$
(corresponding to low $\kappa_L$), we identify 189 compounds. As a
validation, we explicitly compute the lattice dynamical properties for
two of the identified compounds, Cu$_2$AgBiPbS$_4$ and
MnTl$_2$As$_2$S$_5$, and successfully find ultralow $\kappa_L$ values
of $\sim$ 0.3--0.4 W/(m$\cdot$K) at room temperature. In addition to
providing promising ultralow $\kappa_L$ materials, our data-driven
approach opens up new avenues for inverse design of lattice dynamical
properties.


Our design space consists of the materials within the Open Quantum
Materials Database (OQMD),
\cite{saal_materials_2013,kirklin_open_2015} which contains electronic
structure calculations based on density functional theory
(DFT)\cite{hohenberg_inhomogeneous_1964,kohn_self-consistent_1965} for
$\sim$ 635,000 (as of November 2019) known and hypothetical inorganic
crystals derived from the Inorganic Crystal Structure Database
(ICSD)\cite{bergerhoff_inorganic_1983,belsky_new_2002} and structural
prototypes. Our initial screening, before taking into account
$\overline{\alpha}^{(s)}$, is based on several criteria that must be
simultaneously satisfied:
\begin{enumerate}
\item \textbf{Chemical elements} -- Compound must contain one or more
  of the pnictogen group elements that commonly occur with a $3+$
  oxidation state (As, Sb, Bi), one or more chalcogen elements (S, Se,
  Te), and (for practicality) no radioactive elements.
\item \textbf{Electron count} -- The pnictogen must have an oxidation
  state of 3+ (so it is a LP cation) and all other elements must have
  integer oxidation states (to focus on possible
  semiconductors/insulators).
\item \textbf{Crystal structure} -- Each pnictogen atom must be
  coordinated solely by chalcogen atoms and must have a coordination
  number $\ge 3$.
\item \textbf{Synthesizability} -- The compound formation energy must
  be at most 25 meV/atom above that of the thermodynamic ground state
  determined by convex hull analysis (consistent with the overall
  scale of metastability for experimentally observed
  chalcogenides\cite{sun_thermodynamic_2016}) and/or reported
  experimentally. This criterion is chosen to focus on synthesizable
  compounds.
\end{enumerate} The criteria are based on pnictogen group LP cations
(as opposed to others like Pb$^{2+}$ and Tl$^+$) and chalcogens since
the data in work of Skoug and Morelli pertain to such chemistries.
Further details about the screening criteria are included in the
Supporting Information.

\begin{figure*}
  \begin{center}
    \includegraphics[width=1.0\linewidth]{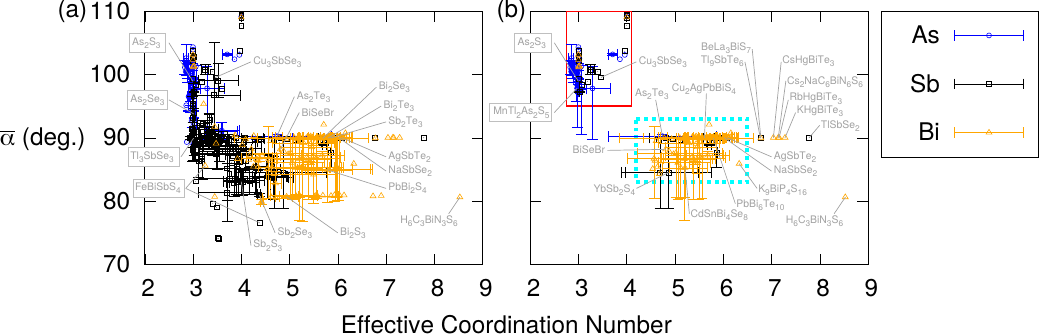}
  \end{center}
  \caption{Coordination number and $\overline{\alpha}^{(s)}$ for each
    pnictogen LP cation element in the (a) 352 compounds passing our
    initial screening criteria and (b) 189 (of the 352) compounds also
    with promising coordination environment, as discussed in the main
    text. Points correspond to average (over LP cation sites) values
    and the ``error bars'' indicate the ranges. A variety of data
    points are labeled with the compound composition to give a general
    sense for the compounds identified. In panel (b), the solid red
    and dotted cyan rectangles show two distinct regions, discussed in
    the main text, containing a vast majority of the compounds.}
  \label{fig:alpha_econ}
\end{figure*}

352 compounds pass the initial screening criteria. Although we have
characterized these compounds in terms of chemical composition,
crystal structure, electronic properties, and synthesizability (see
the Supporting Information), we focus here on the local coordination
environment of the LP cations. Fig. \ref{fig:alpha_econ}(a) shows the
coordination number and $\overline{\alpha}^{(s)}$ of each pnictogen LP
cation element in the 352 compounds. For atomic sites with complex,
low-symmetry coordination environments (often the case for LP
cations), defining coordination number can be challenging. Therefore,
to avoid ambiguity, we employ the well-defined effective coordination
number of Hoppe,\cite{hoppe_effective_1979} a continuous quantity that
includes contributions (smoothly and rapidly decaying with increasing
distance) from all other atoms. For compounds containing multiple
sites of a LP cation element with distinct environments, we show the
average values of coordination number and $\overline{\alpha}^{(s)}$
(which we call $\langle \mathrm{CN} \rangle$ and
$\langle \overline{\alpha} \rangle$, respectively), with the ranges
indicated by the ``error bars.'' The coordination number generally
increases as the LP cation element goes down the pnictogen group of
the periodic table (As $\rightarrow$ Sb $\rightarrow$ Bi). In
addition, consistent with the work of Skoug and Morelli, we generally
find smaller $\overline{\alpha}^{(s)}$ values for larger coordination
number.\cite{skoug_role_2011} Notably, many of the 352 compounds
exhibit large $\overline{\alpha}^{(s)}$ values $\sim 90-100\degree$.

To search for the most promising compounds, we further screen based on
$\overline{\alpha}^{(s)}$. Since the work of Skoug and Morelli
suggests the need for large $\overline{\alpha}^{(s)}$
\textit{relative} to typical values for LP cations of similar
coordination number, we retain a compound if one or more of its
pnictogen LP cation elements is suitable in this sense. Specifically,
for the combination of compound and constituent pnictogen LP cation
element under consideration, its $\langle \overline{\alpha} \rangle$
must be no more than 7\% smaller than the ``best'' value, which we
take to be the maximum value in the following set:
$\langle \overline{\alpha} \rangle$ for all combinations of compound
and constituent LP cation element whose $\langle \mathrm{CN} \rangle$
differs by no more than 0.5 from the one under consideration. For
example, Bi in FeBiSbS$_4$ ($\langle \mathrm{CN} \rangle = 3.25$,
$\langle \overline{\alpha} \rangle = 85.6\degree$) does not satisfy
this criterion since 85.6$\degree$ is more than 7\% smaller than the
largest $\langle \overline{\alpha} \rangle $ value in the
$2.75 \le \langle \overline{\alpha} \rangle \le 3.75$ region of Fig.
\ref{fig:alpha_econ}(a). Since Sb in FeBiSbS$_4$ also does not satisfy
its corresponding criterion, FeBiSbS$_4$ is discarded. The specific
critical value employed (7\%) is fairly arbitrary and is simply chosen
to reduce the number of compounds by around a factor of two.


189 of the 352 compounds, whose coordination number and
$\overline{\alpha}^{(s)}$ are shown in Fig. \ref{fig:alpha_econ}(b),
pass this additional screening criterion. The majority of the
compounds occur in two regions: (1) the red region of smaller
coordination number (containing many As compounds) and (2) the cyan
region of larger coordination number (containing many Bi compounds).
The 189 compounds we have identified should be considered promising
candidates for ultralow $\kappa_L$, and possibly thermoelectric
applications. Indeed, among the 189 are various systems that have been
explored previously for thermoelectricity and exhibit low $\kappa_L$,
such as AgBiS$_2$,\cite{guin_cation_2013}
AgBi$_3$S$_5$,\cite{kim_crystal_2005,tan_high_2017}
Bi$_2$Se$_3$,\cite{hor_p-type_2009,parker_potential_2011,heremans_tetradymites_2017}
K$_2$Bi$_8$Se$_{13}$,\cite{pei_multiple_2016}
AgBiSe$_2$,\cite{morelli_intrinsically_2008,pan_high_2013}
Bi$_2$Te$_3$,\cite{heremans_tetradymites_2017,witting_thermoelectric_2019}
Bi$_2$Te$_2$Se,\cite{tian_understanding_2017,heremans_tetradymites_2017}
AgBiTe$_2$,\cite{irie_thermoelectric_1963,tan_snteagbite2_2014}
GeBi$_2$Te$_4$,\cite{kuznetsova_thermoelectric_2000,shelimova_structural_2000}
GeBi$_4$Te$_7$,\cite{kuznetsov_effect_1999,shelimova_structural_2000,shelimova_crystal_2004}
PbBi$_2$Te$_4$,\cite{kuznetsova_thermoelectric_2000,shelimova_crystal_2004}
PbBi$_4$Te$_7$,\cite{kuznetsova_thermoelectric_2000,shelimova_crystal_2004}
Tl$_9$BiTe$_6$,\cite{wolfing_high_2001,yamanaka_thermoelectric_2003,kurosaki_thermoelectric_2005,guo_enhanced_2013}
Cu$_3$SbSe$_3$,\cite{skoug_role_2011,tyagi_thermoelectric_2014,tyagi_thermoelectric_2014-1}
AgSbTe$_2$,\cite{morelli_intrinsically_2008,wang_synthesis_2008,jovovic_measurements_2008}
TlSbTe$_2$,\cite{kurosaki_thermoelectric_2004,kurosaki_thermoelectric_2005}
and Tl$_9$SbTe$_6$.\cite{guo_enhanced_2013}

In order to validate our approach, we explicitly compute the lattice
thermal transport behavior for two of the identified compounds. We
choose one compound from the cyan region (Cu$_2$AgBiPbS$_4$) and one
from the red region (MnTl$_2$As$_2$S$_5$) of Fig.
\ref{fig:alpha_econ}(b). Both compounds exist (i.e., they are
experimental structures from the ICSD), but they have not been
explored for thermoelectricity. We note that they are metastable
compounds within the OQMD whose formation energies are a small amount
(13--14 meV/atom) above the convex hull.

\begin{figure}[htbp]
  \begin{center}
    \includegraphics[width=1.0\linewidth]{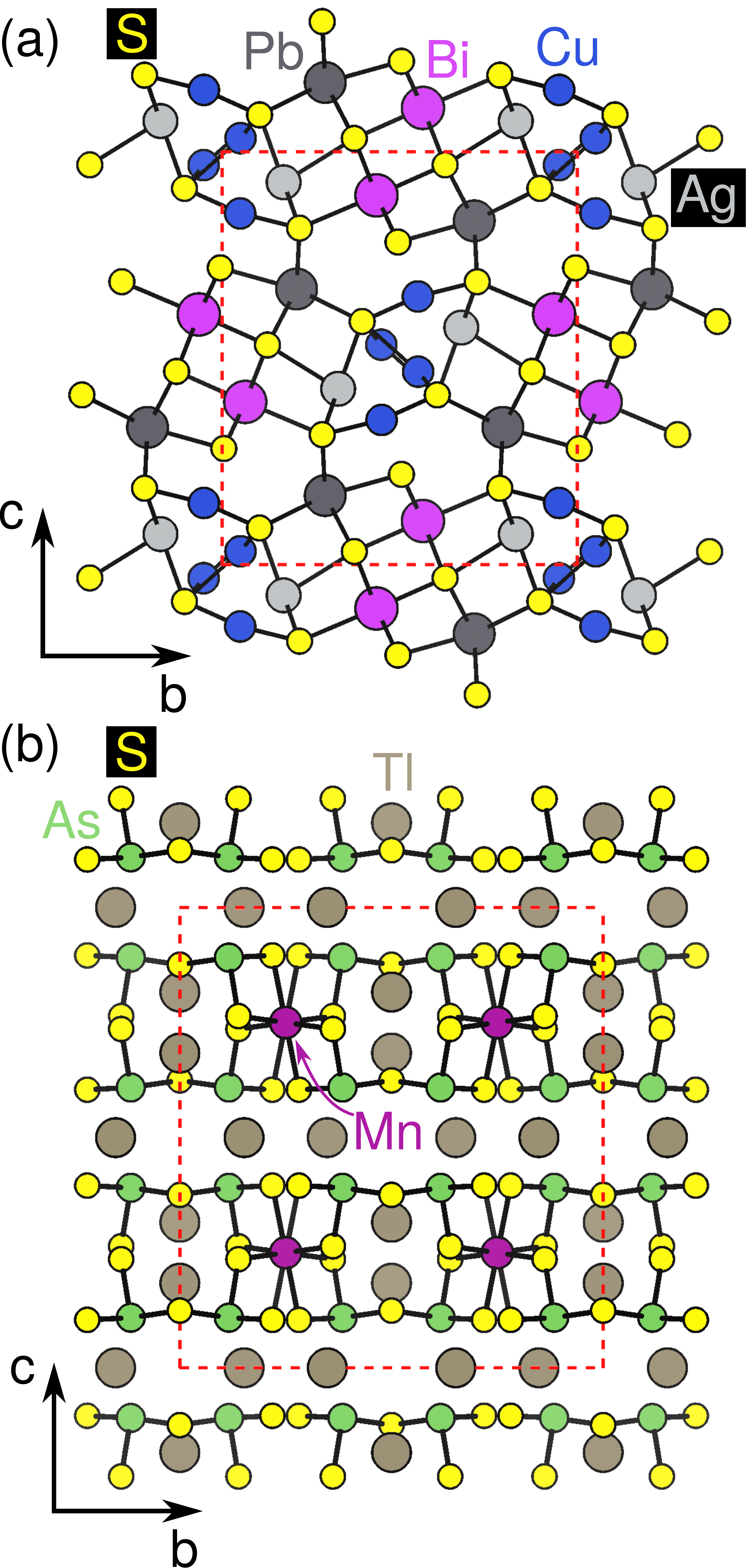}
  \end{center}
  \caption{Crystal structures of (a) Cu$_2$AgBiPbS$_4$ and (b)
    MnTl$_2$As$_2$S$_5$. The unit cell is shown as the dashed red
    line.}
  \label{fig:crystal_structures}
\end{figure}

Fig. \ref{fig:crystal_structures} shows the crystal structures of
Cu$_2$AgBiPbS$_4$ and MnTl$_2$As$_2$S$_5$. Cu$_2$AgBiPbS$_4$, a
mineral
, is orthorhombic ($Pnma$ space group) with a complex 3D bonding
network.\cite{topa_crystal_2010} In Cu$_2$AgBiPbS$_4$, Bi$^{3+}$ is
octahedral with $\langle \overline{\alpha} \rangle=89.98\degree$ and
$\langle \mathrm{CN} \rangle=5.61$, whereas Cu$^+$ exists in linear
and trigonal planar coordinations, Ag$^+$ is trigonal pyramidal, and
Pb$^{2+}$ (also a LP cation) is capped trigonal prismatic.
MnTl$_2$As$_2$S$_5$, which has been synthesized hydrothermally, also
crystallizes with an orthorhombic ($Cmce$) space group, and it can be
considered a layered
structure.\cite{gostojic_crystal_1982,nowacki_crystal_1982} In
MnTl$_2$As$_2$S$_5$, As$^{3+}$ is trigonal pyramidal with
$\langle \overline{\alpha} \rangle=97.82\degree$ and
$\langle \mathrm{CN} \rangle=2.97$, Mn$^{2+}$ is octahedral, and
Tl$^+$ (also a LP cation) is square pyramidal.

\begin{figure*}
  \begin{center}
    \includegraphics[width=1.0\linewidth]{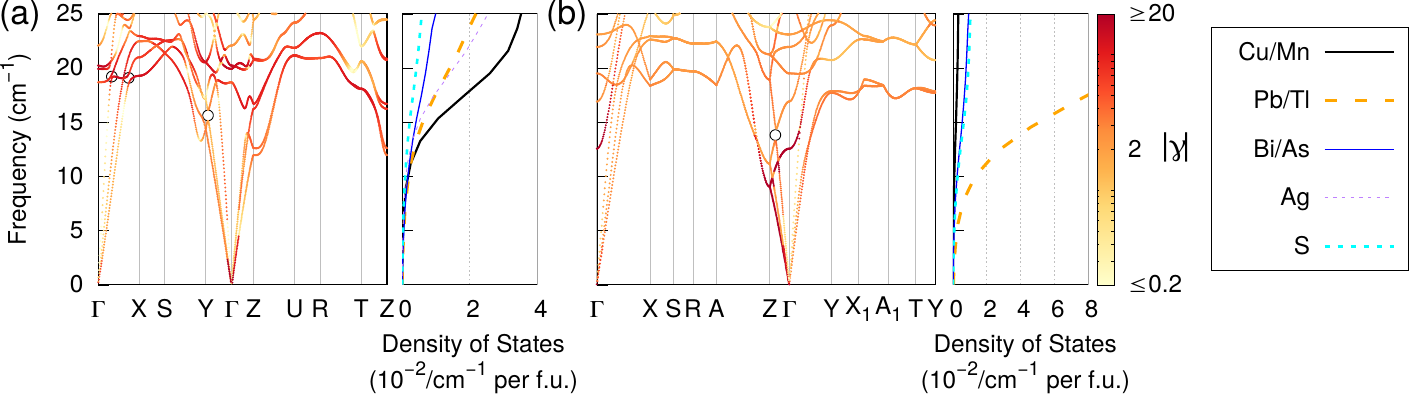}
  \end{center}
  \caption{Low-frequency phonon dispersion and projected phonon
    density of states of (a) Cu$_2$AgBiPbS$_4$ and (b)
    MnTl$_2$As$_2$S$_5$. The line color indicates the magnitude of the
    mode Gr\"{u}neisen parameter $\gamma$. Circles are shown at
    avoided crossings.}
  \label{fig:phonons}
\end{figure*}

\begin{table}[htbp]
  \renewcommand{\arraystretch}{1.2}
  \begin{tabular}{|c|c|c|c|}\hline
    \multicolumn{4}{|c|}{Cu$_2$AgBiPbS$_4$}\\ \hline
    & TA & TA$'$ & LA \\ \hline
    $v_x$ & 1.2 & 1.8 & 3.2 \\
    $v_y$ & 1.3 & 1.4 & 3.0 \\
    $v_z$ & 1.5 & 1.9 & 1.9 \\
    $\Theta_{\Gamma X}$ & 28 & 32 & 30 \\
    $\Theta_{\Gamma Y}$ & 21 & 21 & 25 \\
    $\Theta_{\Gamma Z}$ & 17 & 23 & 18 \\
    $\gamma_{\Gamma X}$ & 6.7 & 6.8 & 10.4 \\
    $\gamma_{\Gamma Y}$ & 3.1 & 1.6 & 5.2 \\
    $\gamma_{\Gamma Z}$ & 2.2 & 4.5 & 3.3 \\ \hline\hline
    \multicolumn{4}{|c|}{MnTl$_2$As$_2$S$_5$}\\ \hline
    & TA & TA$'$ & LA \\ \hline
    $v_x$ & 1.2 & 1.5 & 2.2 \\
    $v_y$ & 1.4 & 1.8 & 2.7 \\
    $v_z$ & 1.1 & 1.3 & 2.2 \\
    $\Theta_{\Gamma X}$ & 33 & 26 & 33 \\
    $\Theta_{\Gamma Y}$ & 26 & 25 & 37 \\
    $\Theta_{\Gamma Z}$ & 13 & 16 & 16 \\
    $\gamma_{\Gamma X}$ & 2.4 & 2.4 & 2.2 \\
    $\gamma_{\Gamma Y}$ & 3.4 & 3.2 & 1.5 \\
    $\gamma_{\Gamma Z}$ & 22.5 & 6.4 & 2.8 \\ \hline
  \end{tabular}
  \caption{Group velocity $v$ at $\Gamma$ (in km/s), Debye temperature
    $\Theta$ (in K), and root-mean-square Gr\"{u}neisen parameter for
    each acoustic branch in each direction.}
  \label{tab:acoustic_properties}
\end{table}

We focus on lattice dynamical properties, but details on the
electronic properties are included in the Supporting Information for
completeness. The low-frequency phonon dispersions for
Cu$_2$AgBiPbS$_4$ and MnTl$_2$As$_2$S$_5$ are shown in Fig.
\ref{fig:phonons}. Due to partial occupancy of the trigonal planar Cu
site (not considered in the OQMD), we employ a unit cell doubled along
the $x$ direction to study Cu$_2$AgBiPbS$_4$, as discussed in detail
in the Supporting Information. Both materials exhibit acoustic and
optical phonons that are quite low in frequency. In the case of
Cu$_2$AgBiPbS$_4$, vibrations of all four cations contribute
appreciably to these modes, with Cu vibrations contributing the most.
For MnTl$_2$As$_2$S$_5$, Tl vibrations are dominant. The computed
sound velocities and Debye temperatures (taken to be the zone boundary
frequencies) for each acoustic branch and direction are shown in Table
\ref{tab:acoustic_properties}. Both exhibit sound velocities of $\sim$
1--3 km/s and, due in part to the complexity of the unit cell, Debye
temperatures of only $\sim$ 15--40 K. Therefore, both
Cu$_2$AgBiPbS$_4$ and MnTl$_2$As$_2$S$_5$ are very soft in terms of
elastic (harmonic) properties.

To measure the anharmonicity in Cu$_2$AgBiPbS$_4$ and
MnTl$_2$As$_2$S$_5$, we compute the mode Gr\"{u}neisen parameter.
Defined as $\gamma =-\frac{\partial\omega/\omega}{\partial V/V},$
where $\omega$ is phonon frequency and $V$ is crystal volume, the
Gr\"{u}neisen parameter provides a measure of the strength of the
phonon--phonon scattering, which limits lattice heat transport. As
shown in Fig. \ref{fig:phonons} and Table
\ref{tab:acoustic_properties}, both compounds exhibit large
Gr\"{u}neisen parameter for the acoustic and especially the low-lying
optical modes. For example, for the optical modes, we find values of
$\sim -15$ for Cu$_2$AgBiPbS$_4$ and $\sim 20$ for
MnTl$_2$As$_2$S$_5$. For the MnTl$_2$As$_2$S$_5$ case, the lowest
optical mode at the zone center contains substantial Tl motion.
Therefore, especially given the large experimental atomic displacement
parameters for Tl (28--36 $\times\ 10^{-3}$
\AA$^2$),\cite{gostojic_crystal_1982} the concept of ``rattler'' Tl
atoms\cite{christensen_avoided_2008,toberer_phonon_2011} may be
relevant to MnTl$_2$As$_2$S$_5$. In contrast, for the
Cu$_2$AgBiPbS$_4$ case, the lowest optical mode at the zone center
corresponds to a collective motion of all atom types. Visualizations
of various phonon modes are included in the Supporting Information.


Determining a highly accurate prediction of $\kappa_L$ for complex
crystals is a significant challenge and is outside the scope of this
work. However, to establish a baseline estimate for the magnitude, we
employ the Debye-Callaway
model,\cite{callaway_model_1959,morelli_estimation_2002} which has
been used to provide a good qualitative picture for a variety of
low-$\kappa_L$
materials.\cite{zhang_first-principles_2012,zhang_first-principles_2016,tan_non-equilibrium_2016,tan_high_2017,zhao_quaternary_2018}
In this approach, for which the data in Table
\ref{tab:acoustic_properties} serve as input, only acoustic phonon
scattering is considered. Due to the soft elastic properties, complex
unit cell, and substantial anharmonicity, the computed Debye-Callaway
model $\kappa_L$ values are only 0.01--0.03 W/(m$\cdot$K) for
Cu$_2$AgBiPbS$_4$ and 0.01--0.07 W/(m$\cdot$K) for
MnTl$_2$As$_2$S$_5$. Such low values, which can only result since the
Debye-Callaway model does not consider that the interatomic distance
is a lower bound to the phonon mean free path, are below the minimum
possible (amorphous limit) $\kappa_L$. Therefore, we take the minimum
$\kappa_L$, via the Cahill model,\cite{cahill_lower_1992} as our best
expectation for the measured lattice thermal conductivity. We find
minimum $\kappa_L$ of 0.44, 0.40, and 0.38 W/m$\cdot$K in the $x$,
$y$, and $z$ directions, respectively, for Cu$_2$AgBiPbS$_4$.
Similarly, we find minimum $\kappa_L$ of 0.31, 0.38, and 0.29
W/m$\cdot$K in the $x$, $y$, and $z$ directions, respectively, for
MnTl$_2$As$_2$S$_5$. Both compounds thus are expected to successfully
achieve the desired ultralow $\kappa_L$, validating our approach.


Our inverse design strategy is successful in that we have identified
materials (fully listed in the Supporting Information) very likely to
exhibit the desired ultralow $\kappa_L$. A natural question is whether
this success stems from solely a correlation between the properties of
the LP and $\kappa_L$, or whether the relationship is causal.
Previously, it has been proposed that the large polarizability of the
LP does cause anharmonicity due to electrostatic interaction between
the LP and the bonding
states,\cite{petrov_heat_1962,skoug_role_2011,zhang_first-principles_2012,nielsen_lone_2013}
As evidence, Zhang \textit{et al.} observed that anharmonic phonon
modes in Cu$_3$SbSe$_3$ involve motion of the Sb$^{3+}$ in the
direction of the LP,\cite{zhang_first-principles_2012} and Nielsen
\textit{et al.} found that the Sb$^{3+}$ LP is encompassed by the
polarization response to Se atomic motion associated with anharmonic
modes in NaSbSe$_2$.\cite{nielsen_lone_2013} In addition, it has been
argued that the further spatial removal of the LP from the LP cation
in particular, as represented by large
$\langle \overline{\alpha} \rangle$, enhances the
anharmonicity.\cite{skoug_role_2011}

Isolating the specific role of the LP in producing low $\kappa_L$
remains a significant challenge.\cite{tolborg_expression_2020} To
investigate this role for Cu$_2$AgBiPbS$_4$ and MnTl$_2$As$_2$S$_5$,
we perform calculations to assess the coupling of the LP to the
anharmonic phonon modes. As discussed in the Supporting Information,
based on our analysis of the electronic localization function and
electronic density of states, we do not find evidence for an
especially strong coupling. In particular, we find no dramatic change
in either the spatial location or the energy of the LP (which is far
below the valence band edge in our case) as a result of the anharmonic
phonons. As such, for the compounds in this study, we are unable to
claim that the presence of the LP, or the specific LP cation
coordination environment or spatial distribution of the LP, directly
causes the ultralow $\kappa_L$. In any case, our data-driven inverse
design approach based on the established correlation between LP cation
properties is capable of providing promising materials with ultralow
$\kappa_L$ and opens the door to future strategies for inverse design
of phononic properties.

\section{Computational Methods}

DFT calculations using a plane-wave basis set and the projector
augmented wave
method\cite{blochl_projector_1994,kresse_ultrasoft_1999,vasp_paw} are
performed using \textsc{vasp}.\cite{kresse_efficiency_1996} We employ
the generalized gradient approximation of Perdew, Burke, and
Ernzerhof,\cite{perdew_generalized_1996} a 500 eV plane wave energy
cutoff, and uniform $k$-meshes with $k$-point density $\ge 700$
$k$-points/\AA$^{-3}$. The energy and ionic forces are converged to
within $10^{-6}$ eV and 0.01 eV/\AA, respectively. Lattice dynamical
properties are computed using \textsc{phonopy}\cite{phonopy} with
approximately cubic
supercells\cite{erhart_first-principles_2015,ase-paper} of 216 atoms
for Cu$_2$AgBiPbS$_4$ and 200 atoms for MnTl$_2$As$_2$S$_5$. We use
atomic displacements of 0.005 \AA\ to compute the phonons and volume
differences of $\pm 1.5$\% to compute the mode Gr\"{u}neisen
parameter. $\kappa_L$ is computed via the Debye-Callaway
model\cite{callaway_model_1959,morelli_estimation_2002,zhang_first-principles_2016}
and compared to the minimum (amorphous limit)
values.\cite{cahill_lower_1992} The reciprocal space high-symmetry
paths are based on Ref. \citenum{setyawan_high-throughput_2010}.


\begin{acknowledgement}
  We acknowledge support from the U.S. Department of Energy under
  Contract DE-SC0014520 (lattice dynamical calculations) and Toyota
  Research Institute through the Accelerated Materials Design and
  Discovery program (materials design). Computational resources were
  provided by the National Energy Research Scientific Computing Center
  (U.S. Department of Energy Contract DE-AC02-05CH11231), the Extreme
  Science and Engineering Discovery Environment (National Science
  Foundation Contract ACI-1548562), and the Quest high performance
  computing facility at Northwestern University.
\end{acknowledgement}

\begin{suppinfo}
  Full list and characterization of the identified compounds,
  screening criteria details, results and discussion of dynamical
  instability in Cu$_2$AgBiPbS$_4$, acoustic branch definitions and
  visualization of phonon modes of Cu$_2$AgBiPbS$_4$ and
  MnTl$_2$As$_2$S$_5$, and calculations of electronic properties and
  coupling with anharmonic phonons of Cu$_2$AgBiPbS$_4$ and
  MnTl$_2$As$_2$S$_5$. 
\end{suppinfo}

\bibliography{lone_pair}

\end{document}